\documentclass[aps,twocolumn,showpacs]{revtex4}
\usepackage{epsfig}
\usepackage{graphicx}
\usepackage{textcomp}

\begin{document}
\title{Multiple extinction routes in stochastic population models}

\author{Omer Gottesman$^{1,2}$ and Baruch Meerson$^{2}$}
\affiliation{$^{1}$Weizmann Institute of Science, Rehovot 76100, Israel}
\affiliation{$^{2}$Racah Institute of Physics, Hebrew University of Jerusalem, Jerusalem 91904, Israel}

\begin{abstract}
Isolated populations ultimately go extinct because of the intrinsic noise of elementary processes. In multi-population systems extinction of a population may occur via more than one route. We investigate this generic situation in a simple predator-prey (or infected-susceptible) model. The predator and prey populations may coexist for a long time but ultimately both go extinct. In the first extinction route the predators go extinct first, whereas the prey thrive for a long time and then also go extinct. In  the second route the prey go extinct first causing a rapid extinction of the predators. Assuming large sub-population sizes in the coexistence state, we compare the probabilities of each of the two extinction routes and predict the most likely path of the sub-populations to extinction. We also suggest an effective three-state master equation for the probabilities to observe the coexistence state, the predator-free state and the empty state.

\end{abstract}

\maketitle

\section{Introduction}
\label{sec:intro}
Isolated populations ultimately go extinct with probability one, even in the absence of adverse environmental variations, because of the presence of an absorbing state at zero population size and the intrinsic noise of elementary processes. Population extinction risk is an important
negative factor in viability of small populations \cite{Bartlett,TREE},
whereas extinction of an endemic disease from a community \cite{Bartlett}
is of course a good thing. For decades, quantitative analysis of population extinction, caused by intrinsic and extrinsic noises, has been in the focus of attention from
bio-mathematicians,  see e.g. Ref. \cite{TREE} for a review. More recently, it has also attracted much attention from physicists \cite{Turner,ElgartPRE70,Doering,AM2006,Frey,Kessler,plasmids,DykmanPRL101,MeersonPRE77,KMS,AKMmod,MeersonPRE79,KamPark,MSepidemic,KhasinPRL103,AssafPRE81,timeresolved,MeersonPRE81,AMS,MS10,KD2,LM2011}, as a highly relevant example of a rare large fluctuation far from thermal equilibrium.

In multi-population systems extinction of a population may occur via more than one route. In this paper we analyze this generic situation in a simple predator-prey model where the predators need the prey for survival. The same model can describe spread of an infectious disease in an isolated community. Although the predator and prey populations (or infected and susceptible populations) may coexist for a long time, they ultimately both undergo extinction, and this happens via one of two routes. In the first route, that can be called sequential,  the predators go extinct first, whereas the prey thrive for a long time and then also go extinct. In  the second route, that can be called (almost) parallel, the prey go extinct first, causing a rapid extinction of the predators. Assuming large sub-population sizes in the coexistence state, we apply a WKB (after Wentzel, Kramers and Brillouin) approximation to the pertinent master equation. In this way we evaluate the probabilities of each of the two extinction routes and predict the most likely path of the sub-populations on the way to extinction. We also suggest an effective three-state master equation for the evolution of the probabilities to observe the coexistence state, the predator-free state and the empty state.

\section{Model and Mean-Field Dynamics}
\label{sec:model}

We assume that predators ($F$ - foxes) and prey ($R$ - rabbits) are well mixed in space so that spatial degrees of freedom are irrelevant. The rabbits reproduce naturally, whereas the foxes only reproduce by predation. The foxes and rabbits die or leave with constant per-capita rates, in general different for each population. We also account, in a model form, for the competition  of rabbits over resources, by adding the elementary process $2R\to R$ that becomes more effective at large population sizes. By taking into account the competition, this model generalizes the classical Lotka-Volterra model \cite{Murray}. The elementary processes and their rates are described in Table \ref{model}. We chose the units of time so that the per-capita death rate of the foxes is equal to 1. The large parameter $N$ determines the typical scaling of the sub-population sizes, see below, and we will assume a strong inequality $N\gg 1$ throughout the paper.

\begin{table}[ht]
\begin{ruledtabular}
\begin{tabular}{|l|c|c|}
 Process & Type of transition &  Rate\\
  \hline
  Reproduction of rabbits & $R\to2R$ & $a$ \\
  Predation/reproduction of foxes & $F+R\to 2F$ & $1/(\Gamma N)$\\
  Death of rabbits & $R\to 0$ &  $b$\\
  Death of foxes & $F\to 0$ & $1$ \\
  Competition among rabbits & $2R\to R$ & $1/N$ \\
\end{tabular}
\end{ruledtabular}
\caption{Predator-prey model}\label{model}
\end{table}

The mean-field equations for this model are
\begin{eqnarray}\label{RFdot}
    \dot{R} &=& (a-b)R-\frac{1}{\Gamma N}RF-\frac{1}{2N}R^2\,,\nonumber \\
    \dot{F} &=& \frac{1}{\Gamma N}RF-F.
\end{eqnarray}
The same model can be reinterpreted to describe the spread of an infectious disease in an \emph{isolated} community.
Here we re-interpret the rabbits as the susceptibles (S) and the foxes as the infected (I). As in the conventional SI model  with population turnover \cite{SI,Grasman,vH,MeersonPRE77,MSepidemic}, a susceptible individual can become infected upon contact with another infected, while infected individuals are removed (recover with immunity, leave or die) with a constant per-capita rate. In the conventional SI model the susceptibles arrive from outside. In the modified SI model, presented here, there are no arrivals from outside, but the susceptibles can reproduce by giving birth. They are also removed (die or leave), as in the conventional SI model. Finally, they compete for resources, $2S\to S$, so their population size remains bounded. See the elementary processes and their rates in Table \ref{model2}, where we measure time in the units of per-capita removal rate of the infected. The mean-field equations for the modified SI model are
\begin{eqnarray}\label{SIdot}
    \dot{S} &=& (a-b)S-\frac{1}{\Gamma N}SI-\frac{1}{2N}S^2\,,\nonumber \\
    \dot{I} &=& \frac{1}{\Gamma N}SI-I.
\end{eqnarray}
These of course coincide with Eqs.~(\ref{RFdot}) upon the change of $S$ to $R$ and $I$ to $F$.
\begin{table}[ht]
\begin{ruledtabular}
\begin{tabular}{|l|c|c|}
 Process & Type of transition &  Rate\\
  \hline
  Reproduction of susceptibles & $S\to2S$ & $a$ \\
  Infection & $I+S\to 2S$ & $1/(\Gamma N)$\\
  Removal of susceptibles & $S\to 0$ &  $b$\\
  Removal of infected & $I\to 0$ & $1$ \\
  Competition among susceptibles & $2S\to S$ & $1/N$ \\
\end{tabular}
\end{ruledtabular}
\caption{Susceptible-Infected model for an isolated community}\label{model2}
\end{table}

\begin{figure}
\includegraphics[width=2.0 in,clip=]{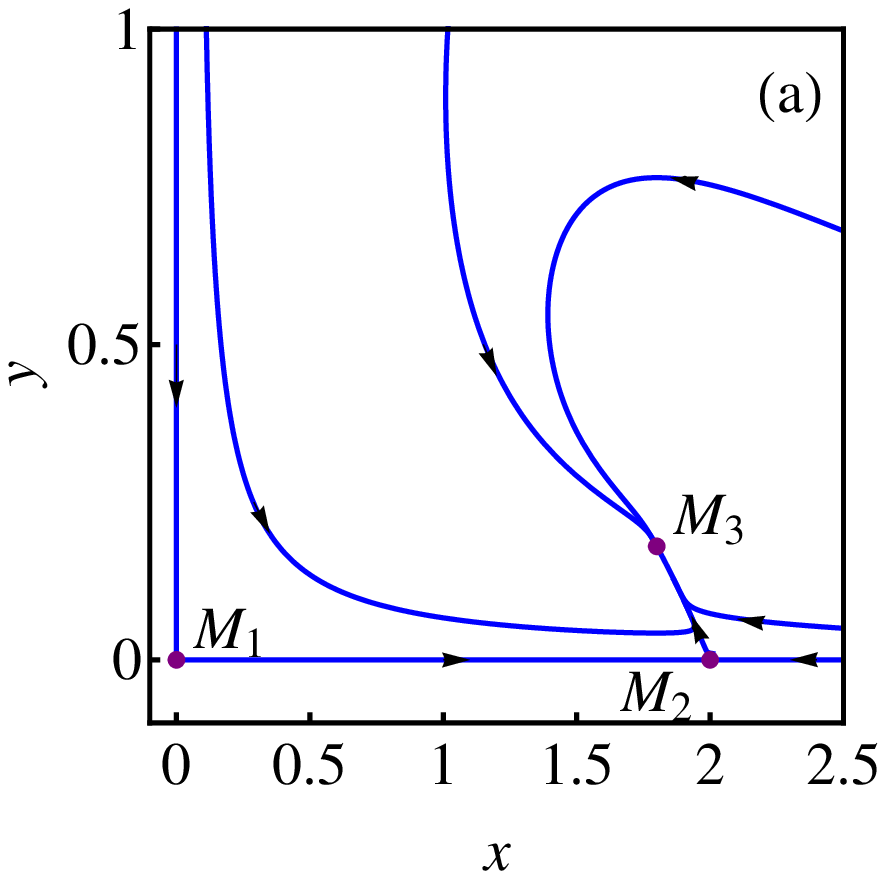}
\includegraphics[width=2.0 in,clip=]{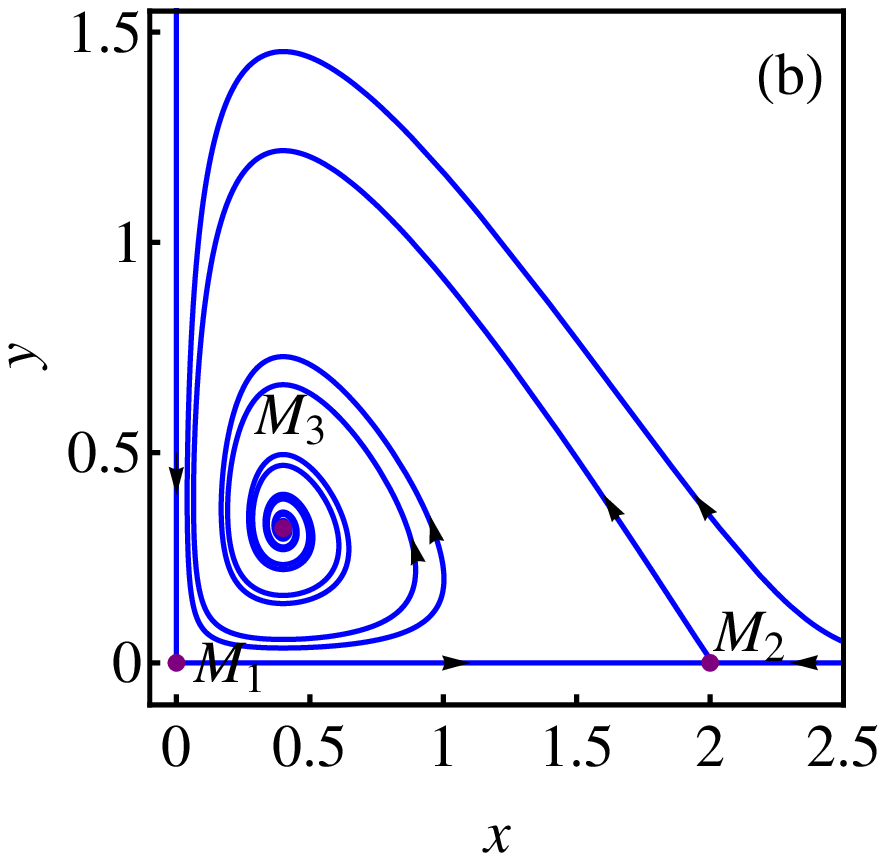}
\caption{(Color online) The mean field phase trajectories of the system for $a=2$, $b=1$ and two different values of $\Gamma$. (a) $\Gamma=1.8$: M3 is a stable node. (b) $\Gamma=0.4$:  M3 is a stable focus.}
\label{MeanFieldPhase}
\end{figure}

For concreteness, we will use the predator-prey notation in the following. Introducing the rescaled population sizes $x=R/N$ and $y=F/N$, we can rewrite the mean-field equations (\ref{RFdot}) as
\begin{equation}\label{xydot}
    \dot{x} = (a-b)x-\frac{xy}{\Gamma}-\frac{x^2}{2}\,,\;\;\;\;\;\dot{y}=\frac{xy}{\Gamma}-y\,.
\end{equation}
The mean-field equations are fully characterized by two parameters, $a-b$ and $\Gamma$. We will be only interested in the case of $a-b>0$ and  $\Gamma<\Gamma_*=2(a-b)$, when Eqs.~(\ref{xydot}) have three fixed points corresponding to non-negative population sizes. The fixed point $M_1$ ($\bar{x}_1=0$, $\bar{y}_1=0$) describes an empty system. It is a saddle point: attracting in the $y$-direction (when there are no rabbits), and repelling in the $x$-direction. The fixed point $M_2$ [$\bar{x}_2=\Gamma_*$, $\bar{y}_2=0$] describes an established population of rabbits in the absence of foxes. It is also a saddle point: attracting in the $x$-direction (when there are no foxes), and repelling in a direction corresponding to the introduction of a small number of foxes into the system. The third fixed point $M_3$ [$\bar{x}_3=\Gamma$, $\bar{y}_3=\Gamma(\Gamma_*-\Gamma)/2$] is attracting and describes the coexistence state. It is a stable node for $\Gamma>\Gamma_0=4(\sqrt{1+a-b}-1)$, and a stable focus for $\Gamma<\Gamma_0$.  Note that $\Gamma_0<\Gamma_*$ for any $a>b$.
Note also that $\bar{y}_3$ is a non-monotonic function of $\Gamma$. It vanishes at  $\Gamma=0$ and $\Gamma=\Gamma_*$, and reaches a maximum, $\Gamma_*^2/8$, at $\Gamma=\Gamma_*/2$ corresponding to the optimal predation rate.  Figure \ref{MeanFieldPhase} shows two examples of mean-field trajectories: for $\Gamma_0<\Gamma<\Gamma_*$ (a) and for $\Gamma<\Gamma_0$ (b). The characteristic relaxation time scale $t_r$ of the coexistence state is determined by the real part of the eigenvalues of the linear stability matrix at $M_3$.

Let us compare the mean-field dynamics of this model with those of the classical Lotka-Volterra model \cite{Murray} and of the SI model with population turnover \cite{SI,MeersonPRE77,MSepidemic}. The competition among the rabbits eliminates one undesirable feature of
the Lotka-Volterra model: the unlimited proliferation of rabbits in the absence of foxes. Furthermore, there are no neutral cycles in this model, in contrast to the Lotka-Volterra model. The attracting fixed point $M_3$, describing a unique coexistence state of the rabbits and foxes, appears instead, with either oscillatory (underdamped) or non-oscillatory (overdamped) character of phase trajectories approaching it. With account of the discreteness of rabbits and foxes,  and of the stochastic character of their interaction, this leads to an exponentially long life-time of the coexistence state, in contrast to a much shorter, power-law life-time, characteristic of the classical Lotka-Volterra model \cite{Frey,KamPark}.

On the other hand, the mean-field dynamics (\ref{SIdot}) is quite similar to that of the conventional SI or SIR model with population turnover \cite{SI,MeersonPRE77,MSepidemic}, except that the fixed point $M_1 (0,0)$, that is absent in the conventional SI model, now appears. As this fixed point is repelling in the $x$-direction, its presence does not make much difference in the \emph{mean-field} description of the coexistence (or, in the context of the SI model, of the endemic state of the disease in the population).

\section{Stochastic analysis}
\label{sec:stochastic}

\subsection{Two extinction routes}
\label{sec:stochastic:two}

The situation, however, changes considerably when one accounts for the stochasticity. Here the empty system is an absorbing state describing extinction of both sub-populations. Having reached this state via a rare large fluctuation, the system will stay there forever. It is this empty state that represents the only true steady state of this system. (This is different from the conventional SI model, where the true steady state describes an established infection-free population.)  Importantly, the absorbing state at zero can typically be reached, at $N\gg 1$, via one of \emph{two} routes. The first route can be called \emph{sequential}: The foxes go extinct first, whereas the rabbits thrive for a long time, and then also go extinct. The presence of this route is caused by the fact that all fox-free states  represent absorbing states for the foxes. The second route is (almost) \emph{parallel}: the rabbits go extinct first causing a rapid (almost deterministic) extinction of the foxes.
\begin{figure}
\includegraphics[width=2.4 in,clip=]{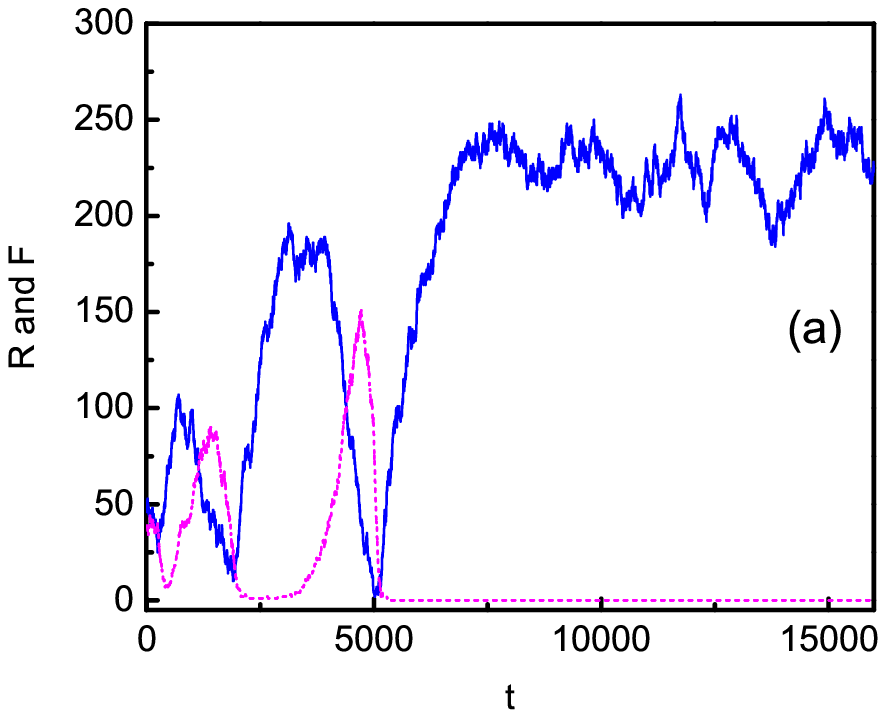}
\includegraphics[width=2.4 in,clip=]{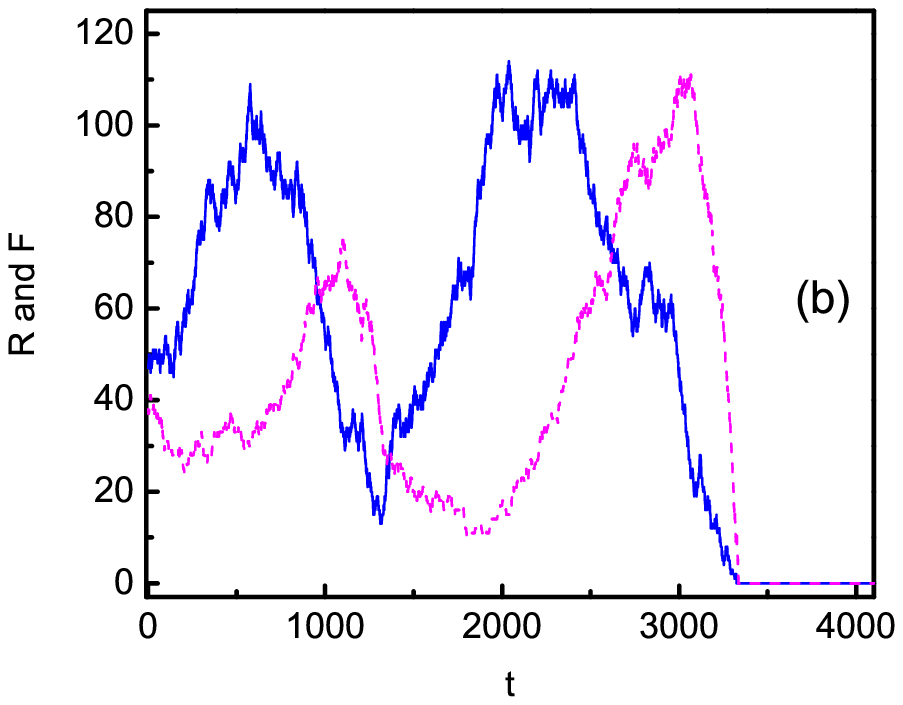}
\caption{(Color online) Two different stochastic realizations of the model for $a=2$, $b=1$, $\Gamma=0.4$ and $N=120$. Shown are the population sizes of rabbits (solid lines) and foxes (dashed lines) versus time (measured in Monte Carlo steps).  (a) The foxes go extinct first, whereas the rabbits get established around the fixed point $M_2$ and, after a very long time, also go extinct (not shown). (b) The rabbits go extinct first, causing a rapid extinction of the foxes.}
\label{MC}
\end{figure}

\begin{figure}
\includegraphics[width=2.4 in,clip=]{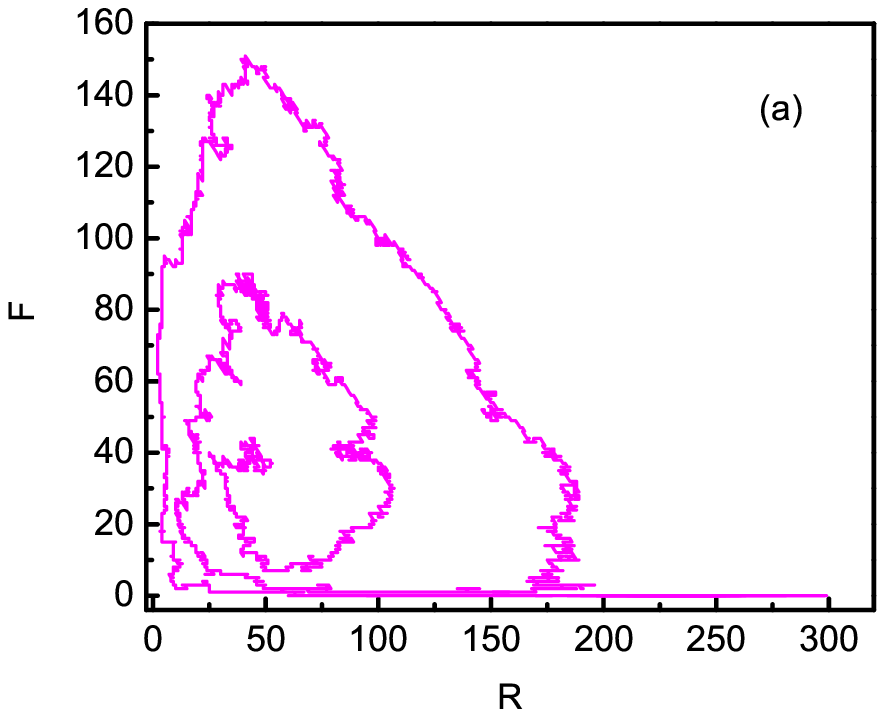}
\includegraphics[width=2.4 in,clip=]{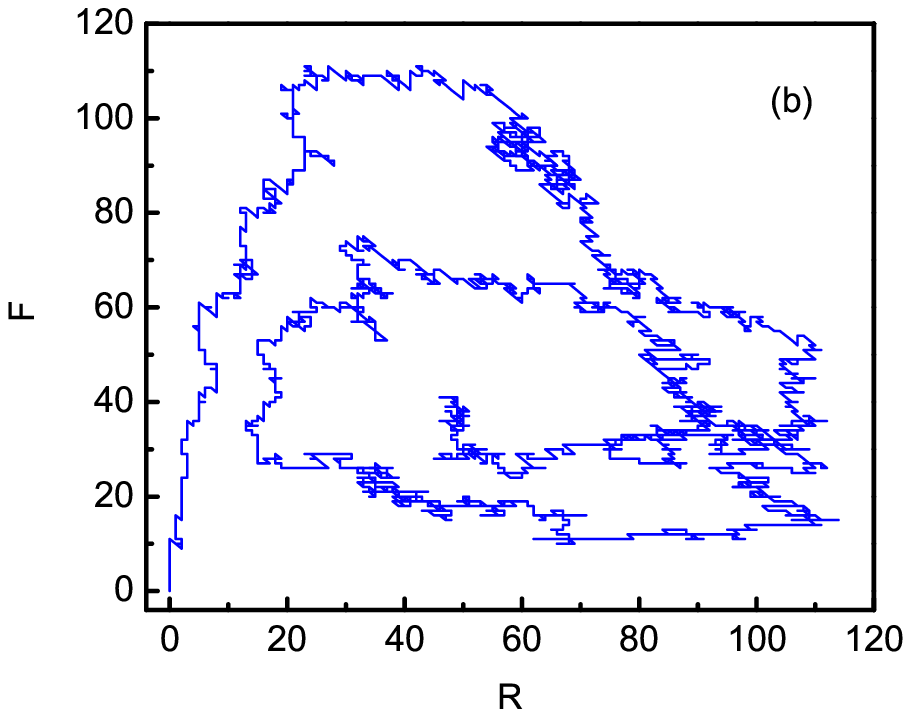}
\caption{(Color online) Same as in Fig. \ref{MC}, but on the $R,F$ plane.}
\label{MCpp}
\end{figure}

The two different routes to extinction are clearly observed from Figure \ref{MC}a and b that shows two different realizations of the stochastic processes, listed in Table 1, for the same set of rate coefficients and for the same initial conditions. The parameters are the same as in Fig. \ref{MeanFieldPhase}b.
In figure a the foxes go extinct first, whereas the rabbit population gets established. Then the rabbits will also go extinct after a very long time (not shown in the figure). In figure b the rabbits go extinct first causing a rapid extinction of the foxes. Figure \ref{MCpp}a and b shows the same trajectories in the $R,F$ plane. One can notice in Figures \ref{MC}b and \ref{MCpp}b that the population of rabbits is very small close to the time when the foxes go extinct. This feature, reproducible in many stochastic simulations,  will obtain a natural explanation in the WKB theory, see subsection C3.
More generally, the WKB theory will enable one to compare the probability of each of the two extinction routes, to predict the most likely path of the rabbits and foxes to extinction, and to evaluate the mean time to extinction of each sub-population.

\subsection{Master equation and long-time dynamics}
\label{sec:stochastic:master}

Let $P_{m,n}(t)$ be the probability to observe, at time $t$, $m$ rabbits and $n$ foxes, where $m,n=0,1,2 \dots$. The evolution of
$P_{m,n}(t)$ is described by the master equation
\begin{eqnarray}
\label{MasterEquation}
    \!\!\dot{P}_{m,n}&\!=\!&\hat{H} P_{m,n}=a[(m-1)P_{m-1,n}-mP_{m,n}]\nonumber \\
    \!\!&\!+\!&(1/\Gamma N)[(m+1)(n-1)P_{m+1,n-1}-mnP_{m,n}]\nonumber \\
    \!\!&\!+\!&b[(m+1)P_{m+1,n}-mP_{m,n}]\nonumber \\
    \!\!&\!+\!&(n+1)P_{m,n+1}-nP_{m,n}\nonumber \\
    \!\!&\!+\!&(1/2N)[(m+1)m P_{m+1,n}-m(m-1)P_{m,n}],
\end{eqnarray}
where $P_{i,j}=0$ when any of the indices is negative. There are three quantities of interest here describing extinction of the sub-populations involved. The probability of extinction, by time $t$, of foxes at any
number of rabbits is equal to $\sum_{m=0}^{\infty} P_{m,0}(t)$. The evolution of this quantity in time is described by the equation
\begin{equation}\label{S1}
    \frac{d}{dt} \sum_{m=0}^{\infty} P_{m,0}(t)=\sum_{m=0}^{\infty} P_{m,1}(t),
\end{equation}
directly following from Eq.~(\ref{MasterEquation}). The right-hand side of Eq.~(\ref{S1}) is simply the probability of death of the last remaining fox at any number of rabbits.

In its turn, $\sum_{n=0}^{\infty} P_{0,n}(t)$ is the probability of extinction, by time $t$, of rabbits at any number of foxes. This quantity evolves in time according to
\begin{equation}\label{S2}
    \frac{d}{dt} \sum_{n=0}^{\infty} P_{0,n}(t)=\frac{1}{\Gamma N}\sum_{n=1}^{\infty} n P_{1,n}(t)+b\sum_{n=0}^{\infty}P_{1,n}(t).
\end{equation}
The first and second terms on the right are the probabilities of predation and of natural death of the last remaining rabbit, respectively. Finally,  the probability $P_{0,0}(t)$ of extinction of both rabbits and foxes by time $t$ is described by the equation
\begin{equation}\label{dP2_over_dt}
    \frac{dP_{0,0}}{dt}=bP_{1,0}(t)+P_{0,1}(t)\,.
\end{equation}
At $N\gg 1$, and at times much longer than the relaxation time $t_r$ of the mean-field theory, the quantities that appear in Eqs.~(\ref{S1})-(\ref{dP2_over_dt}) become sharply peaked around the corresponding fixed points of the mean-field theory. The structure of these peaks is affected by the presence or absence of absorbing states. At long times, the probability distribution $P_{m,0}(t)$ with $m>0$ (which describes the dynamics of rabbits conditional on prior extinction of the foxes) is a one-dimensional distribution peaked at the fixed point $M_2$ of the mean-field theory. The probability distribution  $P_{m,n}(t)$ with
$m,n>0$ (which describes the long-time dynamics of the coexisting rabbits and foxes) is a two-dimensional distribution peaked
at the fixed point $M_3$ of the mean-field theory. Finally, the extinction probability of both sub-populations $P_{0,0}(t)$ corresponds to a Kronecker-delta probability density.  Not only the  structure, but the long-time dynamics of these three distributions are different. To clearly see this point, let us define the total ``probability
contents" of the vicinities of each of the fixed points $M_1$, $M_2$ and $M_3$:
\begin{eqnarray}
  \mathcal{P}_1 (t) &=& P_{0,0} (t)\,, \label{calP1} \\
  \mathcal{P}_2 (t) &=& \sum_{m=1}^{\infty}P_{m,0} (t)\,, \label{calP2} \\
  \mathcal{P}_3 (t) &=& \sum_{m=1}^{\infty} \sum_{n=1}^{\infty} P_{m,n} (t)\,. \label{calP3}
\end{eqnarray}
At $N\gg 1$ and $t\gg t_r$ the sums in Eqs.~(\ref{calP2}) and (\ref{calP3}) are mostly contributed to by close vicinities of $M_2$ and $M_3$, respectively. The long-time evolution of $\mathcal{P}_1$,  $\mathcal{P}_2$ and  $\mathcal{P}_3$ is described by an effective three-state master equation:
\begin{eqnarray}
\label{efrateeqn}
  \dot{\mathcal{P}}_1 (t) &=& r_{21}  \mathcal{P}_2 (t) + r_{31} \mathcal{P}_3 (t)\,,\nonumber \\
  \dot{\mathcal{P}}_2 (t)&=&  -r_{21}  \mathcal{P}_2 (t) + r_{32} \mathcal{P}_3 (t)\,,\nonumber \\
  \dot{\mathcal{P}}_3 (t)&=& -(r_{31} + r_{32}) \mathcal{P}_3 (t) \,,
\end{eqnarray}
where $r_{ij}$ is the (yet unknown) transition rate from the vicinity of the fixed point $i$ to the vicinity of the fixed point $j$. Let the initial condition correspond to the coexistence state around $M_3$:
\begin{equation}\label{incond}
   [\mathcal{P}_1 (0),\,\mathcal{P}_2 (0),\,\mathcal{P}_3 (0)]=(0,0,1)\,.
\end{equation}
Then the solution of Eqs. ~(\ref{efrateeqn}) is
\begin{eqnarray}
\mathcal{P}_1 (t) &=& 1+\frac{r_{32}\,e^{-r_{21}t}+(r_{31}-r_{21})\,e^{-(r_{31}+r_{32})t}}{r_{21}-r_{31}-r_{32}},\label{P1}\\
\mathcal{P}_2 (t) &=&\frac{r_{32}\,[e^{-(r_{31}+r_{32})t}-e^{-r_{21}t}]}{r_{21}-r_{31}-r_{32}},  \label{P2} \\
\mathcal{P}_3 (t)  &=& e^{-(r_{31}+r_{32})t}.\label{P3}
\end{eqnarray}
Once the transition rates $r_{31}$, $r_{32}$ and $r_{21}$ are known, Eqs.~(\ref{P1})-(\ref{P3}) provide a valuable ``coarse-grained" description of the stochastic system in terms of the long-time
evolution of the probabilities to observe the coexistence state around $M_3$, the fox-free state around $M_2$ and the extinction state at $M_1$.  The coexistence probability $\mathcal{P}_3 (t)$ goes down to zero exponentially in time. The  probability $\mathcal{P}_1 (t)$ of extinction of both sub-populations increases
monotonically with time, exhibiting two different exponents, and ultimately reaches $1$. Finally, the probability of the fox-free state
$\mathcal{P}_2 (t)$ first increases with time, reaches a maximum and then goes down to zero. At intermediate times, $t_r\ll t\ll \mbox{min}\, [1/r_{21}, 1/(r_{31}+r_{32})]$, Eqs.~(\ref{P1})-(\ref{P3}) read
\begin{eqnarray}
\mathcal{P}_1 (t) &\simeq & r_{31}t,\;\;\;\;\;\; \mathcal{P}_2 (t) \simeq r_{32}t,\nonumber \\
\mathcal{P}_3 (t)  &\simeq&1-(r_{31}+r_{32})t.\label{Pshort}
\end{eqnarray}
During this stage of the dynamics, $\mathcal{P}_1(t)$ and $\mathcal{P}_2(t)$ grow linearly with time, whereas the transition $2\to 1$ does not show up. What happens at longer times? The transition rates are usually widely different in magnitude. According to our WKB calculations, presented below,
$r_{21}\ll r_{31}+r_{32}$. Then, at times $t\gg  1/(r_{31}+r_{32})$, Eqs.~(\ref{P1})-(\ref{P3}) become
\begin{eqnarray}
\mathcal{P}_1 (t) &\simeq & 1-\frac{r_{32} e^{-r_{21}t}}{r_{31}+r_{32}},\label{P1long}\\
\mathcal{P}_2 (t) &\simeq &\frac{r_{32} e^{-r_{21}t}}{r_{31}+r_{32}},  \label{P2long} \\
\mathcal{P}_3 (t)  &\simeq & 0.\label{P3long}
\end{eqnarray}
Now there is a probability current from state 2 to state 1, describing extinction of rabbits in the absence of foxes, but the transition rates $r_{31}$ and  $r_{32}$ are still present in the equations.

It is common, see e.g. Ref. \cite{TREE},  to characterize the extinction risk of a stochastic population in terms of its mean time to extinction. From Eqs.~(\ref{P1})-(\ref{P3}), the mean time to extinction of foxes is
\begin{eqnarray}
\label{MTEfox}
  \tau_F&=& \int_0^{\infty} dt\,t [\dot{P}_1(t)+\dot{P}_2(t)]= -\int_0^{\infty} dt\,t \dot{P}_3(t)\nonumber \\
  &=&\frac{1}{r_{31}+r_{32}},
\end{eqnarray}
whereas the mean time to extinction of both sub-populations is
\begin{equation}\label{MTEtotal}
    \tau_{RF}=\int_0^{\infty} dt\,t \dot{P}_1(t)=\frac{r_{21}+r_{32}}{r_{21}(r_{31}+r_{32})}.
\end{equation}

How is this dynamics encoded in the spectral properties of the the linear operator $\hat{H}$, see Eq.~(\ref{MasterEquation})? Let $\lambda_i$ be the eigenvalues and $\pi_{m,n}^{(i)}$ the eigenstates of $\hat{H}$. These are defined by the relations
$$
\hat{H} \pi_{m,n}^{(i)} = -\lambda_i  \pi_{m,n}^{(i)},\;\;\;i=1,2,\dots\,,
$$
so that
$$
P_{m,n}(t)=\sum_{i=1}^{\infty}C_i \pi_{m,n}^{(i)} e^{-\lambda_i t}\,,
$$
where the constants $C_i$ are determined by the initial condition $P_{m,n}(0)$. Let us order the eigenvalues so that $\lambda_1< \lambda_2<\lambda_3 <\dots$.  The true (empty) steady state corresponds to the only zero eigenvalue: $\lambda_1=0$, whereas $\pi_{m,n}^{(1)}=\delta_{m 0} \delta_{n 0}$. 
Equations~(\ref{P1})-(\ref{P3}) and the inequality $r_{21}< r_{31}+r_{32}$ imply that, at $N\gg 1$, the next two eigenvalues are $\lambda_2=r_{21}$ and $\lambda_3=r_{31}+r_{32}$. The quantity $r_{21}$ was found in Refs. \cite{MeersonPRE79,AssafPRE81}, and it is exponentially small in $N$:
\begin{equation}\label{r21}
r_{21}\simeq\sqrt{\frac{b N}{\pi}}\frac{(a-b)^2}{a}\,
\exp\left[-2 N\left(a-b+b\,\ln\frac{b}{a}\right)\right].
\end{equation}
The corresponding
eigenvector $\pi_{m,0} \delta_{n0}$ is the quasi-stationary distribution of the \emph{one-population} model $R\to 2R$, $R\to 0$ and $2R\to R$, where the predators are absent \cite{AssafPRE81}. The eigenvalue $r_{31}+r_{32}$ is also exponentially small in $N$, see below. It corresponds to the quasi-stationary distribution $\pi_{m,n}$ at $n>0$, sharply peaked at the fixed point $M_3$. Our main effort in the following will be to evaluate, with exponential accuracy, the transition rates $r_{31}$ and $r_{32}$ that contribute to the eigenvalue  $r_{31}+r_{32}$.
As all the effective transition rates $r_{ij}$ are exponentially small in $N\gg 1$, we can drop the right-hand side in the eigenvalue problem
\begin{equation}\label{QSD_eq0}
    \hat{H}\pi_{m,n}=-(r_{31}+r_{32}) \pi_{m,n}\,,\;\;\;n>0
\end{equation}
and arrive at the quasi-stationary equation
\begin{equation}\label{QSD_eq}
    \hat{H}\pi_{m,n}\simeq 0\,,\;\;\;m>0,\;n>0\,.
\end{equation}

\subsection{WKB approximation}
\label{sec:stochastic:WKB}

\subsubsection{General}

For $N\gg 1$, Eq.~(\ref{QSD_eq}) can be approximately solved via the WKB ansatz
\begin{equation}\label{WKB_ansatz}
    \pi_{m,n}=\exp[-N S(x,y)]\,,
\end{equation}
where $S$ is assumed to be a smooth function of the continuous variables $x=m/N$ and $y=n/N$. The use of WKB approximation for finding stationary or quasi-stationary solutions of master equations with a discrete state space was pioneered in Refs. \cite{Kubo,Hu,Peters,DykmanPRE100}. By now this approximation, in the space of population sizes  or in the momentum space, has become a standard tool
in the analysis of rare large fluctuations of stochastic populations \cite{TREE,MSepidemic,ElgartPRE70,Kessler,DykmanPRL101,MeersonPRE77,AKMmod,MSexplosion,MeersonPRE79,
Escudero,KhasinPRL103,MeersonPRE81,AMS,MS10,KD2,Kaplan,MS11,LM2011,Assaf2011}. Plugging Eq.~(\ref{WKB_ansatz}) into Eq.~(\ref{QSD_eq}) and Taylor expanding $S$ to first order around $(x,y)$, we arrive at a zero-energy Hamilton-Jacobi equation $H(x,y,\partial_xS,\partial_yS)=0$, with the Hamiltonian
\begin{eqnarray}\label{Hamiltonian}
   \!\!\!\!\!\!\!\!&&H(x,y,p_x,p_y)=ax(e^{p_x}-1)+bx(e^{-p_x}-1)\nonumber\\
   \!\!\!\!\!\!\!\!&&+\frac{xy}{\Gamma}(e^{p_y-p_x}-1)+y(e^{-p_y}-1)+\frac{x^2}{2}(e^{-p_x}-1).
\end{eqnarray}
The trajectories are given by the Hamilton's equations for the ``coordinates" $x$ and $y$ and conjugate ``momenta"
$p_x$ and $p_y$:
\begin{eqnarray}\label{Motion}
    \dot{x}&=&axe^{p_x}-\left(bx+\frac{x^2}{2}\right)e^{-p_x}-\frac{xy}{\Gamma}e^{p_y-p_x},\nonumber \\
    \dot{y}&=&\frac{xy}{\Gamma}e^{p_y-p_x}-ye^{-p_y},\nonumber \\
    \dot{p_x}&=&(b+x)(1-e^{-p_x})+\frac{y}{\Gamma}\left(1-e^{p_y-p_x}\right)+a(1-e^{p_x}),\nonumber \\
    \dot{p_y}&=&1-e^{-p_y}+\frac{x}{\Gamma}\left(1-e^{p_y-p_x}\right),
\end{eqnarray}
where we are only interested in zero-energy trajectories. The (zero-energy) invariant hyperplane $p_x=p_y=0$ corresponds to the mean-field dynamics~(\ref{xydot}). The invariant hyperplanes $x=0$ and $y=0$ correspond to the rabbit-free and fox-free dynamics, respectively. The Hamiltonian problem is characterized by three independent parameters $a$, $b$ and $\Gamma$.

The Hamiltonian flow (\ref{Motion}) has five zero-energy fixed points:
\begin{eqnarray}\label{FixedPoints}
    M_1&=&(0,0,0,0),\nonumber \\
    M_2&=&(\Gamma_*,0,0,0),\nonumber \\
    M_3&=&[\Gamma,\Gamma (\Gamma_*-\Gamma)/2,0,0],\nonumber \\
    F_1&=&[0,0,\ln(b/a),0],\nonumber \\
    F_2&=&[\Gamma_*,0,0,-\ln(\Gamma_*/\Gamma)]\,.
\end{eqnarray}
The zero-momentum fixed points $M_1$, $M_2$ and $M_3$ correspond to the three fixed points of the mean field equations (\ref{xydot}), so we keep the same notation for them. The two other fixed points, $F_1$ and $F_2$, are fluctuational fixed points describing a fox-free state at a non-zero number of rabbits ($F_2$) and an empty system ($F_1$). Fluctuational fixed points have a non-zero $p_x$ or $p_y$ component and appear in a broad class of stochastic population models exhibiting extinction in the absence of an Allee effect \cite{TREE,MeersonPRE79,AssafPRE81,Grasman,EK2,DykmanPRL101,MeersonPRE77,MeersonPRE81,KhasinPRL103,MS10,LM2011}.
They play an important role in the calculations of the quasi-stationary distributions and the mean time to extinction. It is mostly their presence that makes the WKB theory of  population extinction distinct from the WKB theory of noise-induced switches between different states that are stable in the mean-field theory \cite{FW}.

To determine $S(x,y)$ in Eq.~(\ref{WKB_ansatz}), one should calculate the action accumulated along the (zero-energy) \emph{activation
trajectory} in the phase space of the Hamilton's equations (\ref{Motion}) that exits 
the fixed point $M_3$ and ends
at $(x,y)$:
$$
S(x,y)=\int_{M_3}^{(x,y)} p_x dx + p_y dy\,,
$$
and then minimize the result with respect to $p_x$ and $p_y$ at the end point $(x,y)$. To evaluate the transition rates $r_{31}$ and $r_{32}$, we need to evaluate $\pi_{m,n}$ at points $(x=0,y=0)$ and $(x=\Gamma_*,y=0)$, respectively. As in many other stochastic population models, there are no phase trajectories that would start at $M_3$ and end at fixed points $M_1$ or $M_2$. There are, however, instanton-like activation trajectories that start, at $t=-\infty$, at $M_3$ and enter,  at $t=\infty$, the fixed point $F_1$ or $F_2$, respectively.
The accumulated actions $S_{31}$ and $S_{32}$ along these instantons,
$$
S_{31}=\int_{M_3}^{F_1} p_x dx + p_y dy\;\;\mbox{and}\;\;S_{32}=\int_{M_3}^{F_2} p_x dx + p_y dy,
$$
yield, with exponential accuracy, the transitions rates $r_{31}$ and $r_{32}$:
\begin{eqnarray}\label{rates}
  r_{31} &\sim & \exp(-NS_{31}) \nonumber \\
  r_{32} &\sim & \exp(-NS_{32})\,.
\end{eqnarray}
The one-dimensional quasi-stationary distribution $\pi_{m,0} \delta_{n,0}$, corresponding to the
long-lived population of rabbits in the absence of foxes, corresponds to another instanton-like trajectory: the one
connecting the fixed points $M_2$ and $F_1$ \cite{MeersonPRE79,AssafPRE81}. The accumulated action
\begin{eqnarray}\label{S21}
    S_{21}=\int_{M_2}^{F_1} p_x dx =2\left(a-b+b\ln\frac{b}{a}\right)
\end{eqnarray}
yields the transition rate $r_{21}$ which agrees, up to a pre-exponential factor, with
Eq.~(\ref{r21}). See Refs. \cite{MeersonPRE79,AssafPRE81} for detail.

\subsubsection{Analytic result for $S_{32}$ near bifurcation $\Gamma=\Gamma_*$}
\label{sec:stochastic:instantons:analytical}
Because of the lack of an independent integral of motion in addition to the Hamiltonian, the canonical equations (\ref{Motion}) are non-integrable analytically for a general set of parameters $a$, $b$ and $\Gamma$. Close to the bifurcation $\Gamma=\Gamma_*$, however, the fixed points $M_3$ and  $F_2$ become very close to each other, and the action $S_{32}$ can be calculated perturbatively by exploiting the time separation property uncovered in Refs. \cite{DykmanPRL101,MeersonPRE77}.
Let us define the bifurcation parameter $\delta=1-\Gamma/\Gamma_*$. For $\delta\ll1$, the fixed points $M_3$ and $F_2$ become $M_3\simeq\left[\Gamma_*(1-\delta),\delta\Gamma_*^2/2,0,0\right]$ and $F_2\simeq\left(\Gamma_*,0,0,-\delta\right)$.
Motivated by these expressions, we assume the following scalings: $x=\Gamma_*+\delta\, q_1$, $y=\delta\Gamma_*^2 q_2/2$, $p_x=\delta^2 p_1$, and $p_y=\delta p_2$, where $q_1$, $q_2$, $p_1$ and $p_2$ are (non-canonical) rescaled variables. The equations of motion become, in the leading order in $\delta$,

\begin{eqnarray}\label{RescaledMotionEq}
    \dot{q}_1&=&-\frac{\Gamma_*}{2}\left(q_1+\Gamma_*q_2\right),\nonumber \\
    \dot{q}_2&=&\delta q_2\left(1+2p_2+\frac{q_1}{\Gamma_*}\right),\nonumber \\
    \dot{p}_1&=&\frac{\Gamma_*}{2}\left(p_1-q_2p_2\right),\nonumber \\
    \dot{p}_2&=&\delta\left(p_1-\frac{q_1p_2}{\Gamma_*}-p_2-p_2^2\right).
\end{eqnarray}
The dynamics of $q_1$ and $p_1$ is fast compared to that of $q_2$ and $p_2$. As a result, $q_1$ and $p_1$ adiabatically follow the slowly varying  $q_2$ and $p_2$: $q_1=-\Gamma_*q_2$, $p_1=q_2p_2$. Plugging these expressions in the equations for $q_2$ and $p_2$, we obtain equations
\begin{eqnarray}\label{SlowSystemMotionEq}
    \dot{q}_2=\delta q_2\left(1+2p_2-q_2\right),\;\;
    \dot{p}_2=\delta p_2\left(2q_2-1-p_2\right),
\end{eqnarray}
derivable from the reduced ``universal" Hamiltonian $H_r(q_2,p_2)=\delta q_2 p_2(p_2-q_2+1)$
\cite{MeersonPRE79,DykmanPRL101,MeersonPRE77,EK2}. This problem is easily soluble, and one obtains
\begin{eqnarray}\label{S32Analytic}
    S_{32}\simeq\frac{\delta^2\Gamma_*^2}{2}\int^0_1(q_2-1)dq_2
    = \frac{(\Gamma_*-\Gamma)^2}{4}.
\end{eqnarray}

\subsubsection{Numerical calculations of $S_{31}$ and $S_{32}$}
\label{sec:stochastic:instantons:numerical}

\begin{figure}
\includegraphics[width=2.7 in,clip=]{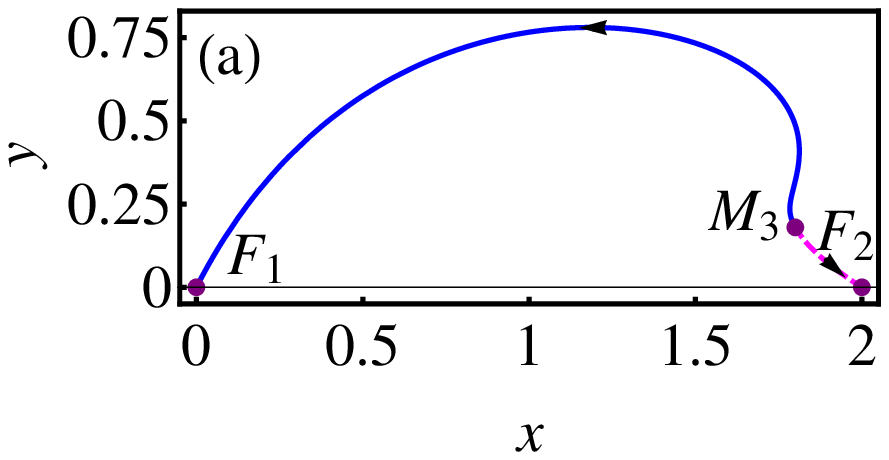}
\includegraphics[width=2.7 in,clip=]{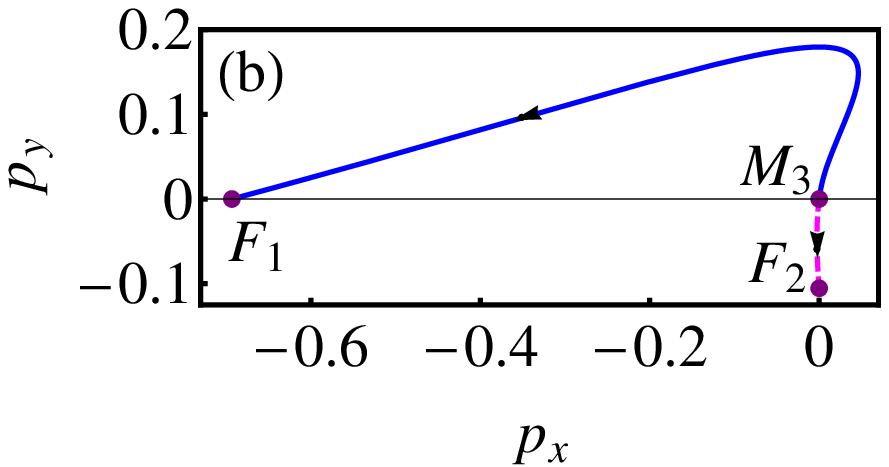}
\caption{The numerically found instantons $M_3F_1$ and $M_3F_2$ for $a=2,\;\;b=1$ and $\Gamma=1.8$, when the fixed point $M_3$ is a node. Shown are
the $xy$-projections (a) and the $p_xp_y$ projections (b).}
\label{nodeinst}
\end{figure}

\begin{figure}
\includegraphics[width=2.7 in,clip=]{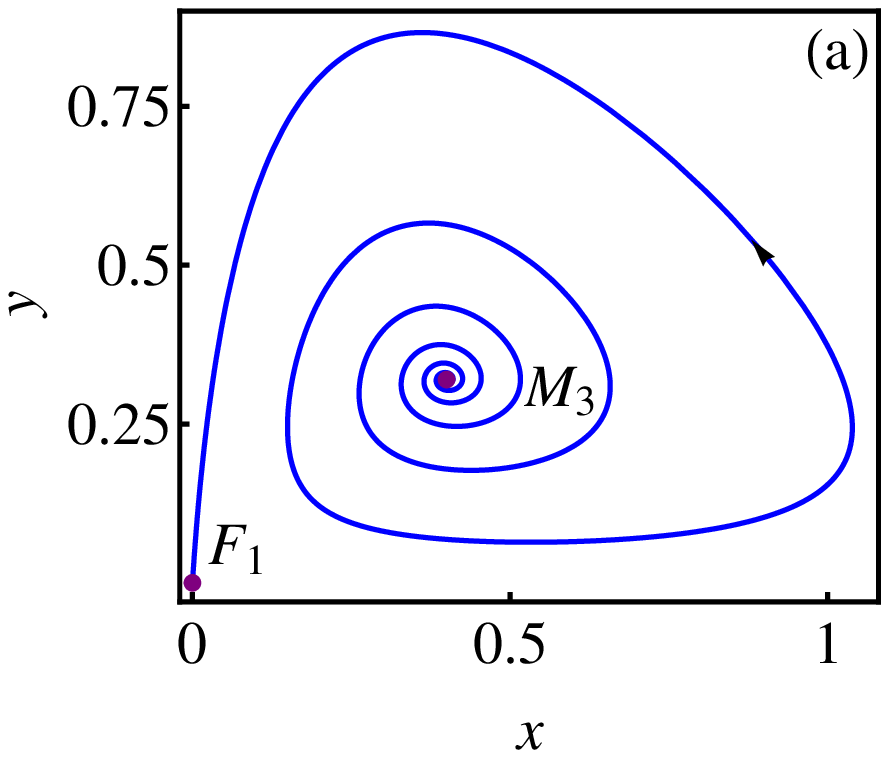}
\includegraphics[width=2.7 in,clip=]{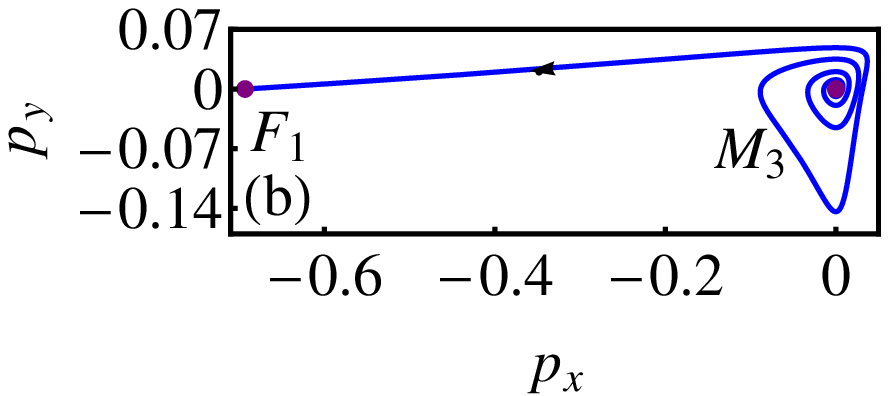}
\caption{The numerically found instanton $M_3F_1$ for $a=2,\;\;b=1$ and $\Gamma=0.4$, when the fixed point $M_3$ is a focus. Shown are
the $xy$-projection (a) and the $p_xp_y$ projection (b).}
\label{focinst31}
\end{figure}

\begin{figure}
\includegraphics[width=2.7 in,clip=]{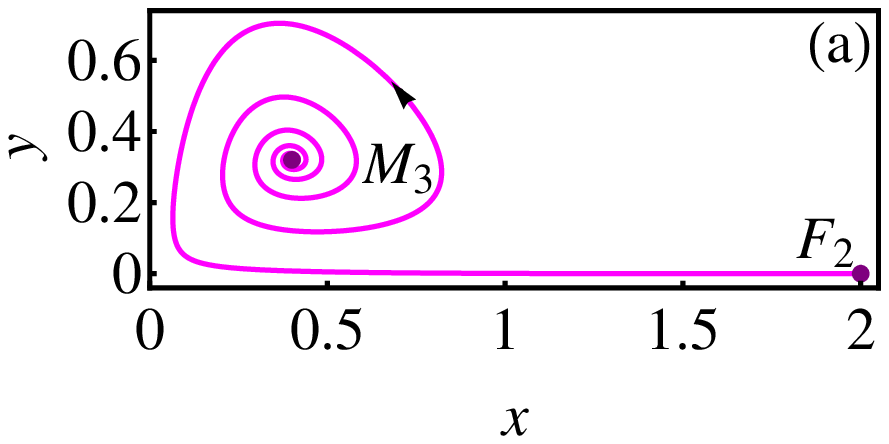}
\includegraphics[width=2.7 in,clip=]{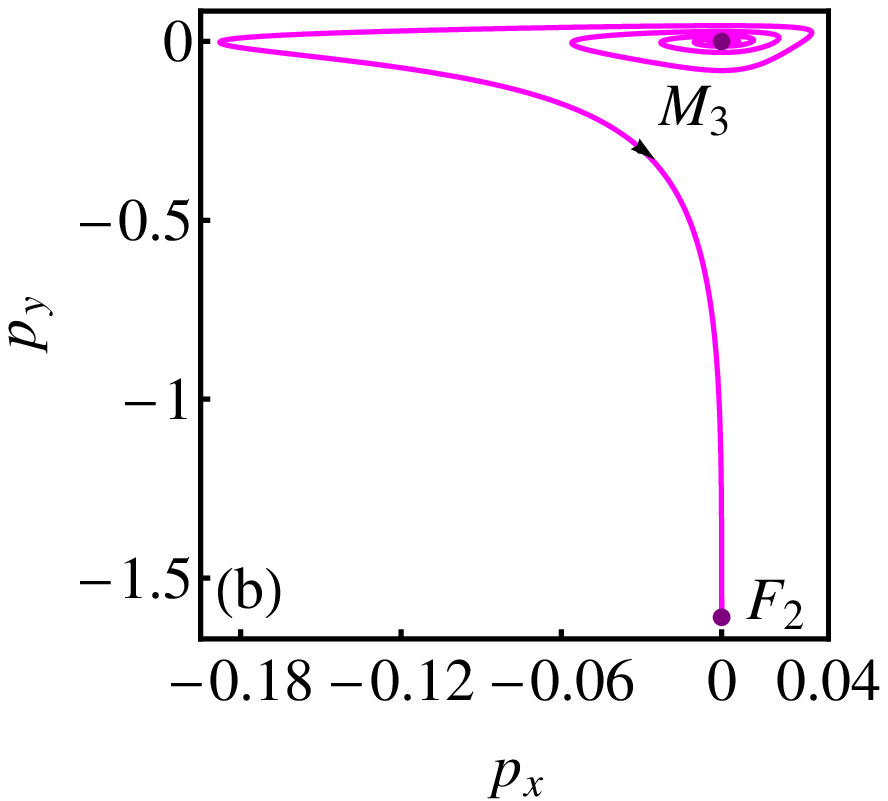}
\caption{The numerically found instanton $M_3F_2$ for $a=2,\;\;b=1$ and $\Gamma=0.4$, when the fixed point $M_3$ is a focus. Shown are
the $xy$-projection (a) and the $p_xp_y$ projection (b).}
\label{focinst32}
\end{figure}

For a general set of parameters $a$, $b$ and $\Gamma$, one can find the instantons $M_3F_1$ and $M_3F_2$ numerically: either by shooting \cite{MeersonPRE77,DykmanPRL101}, or by iterating the equations for $\dot{x}$ and $\dot{y}$ forward in time, and equations for $\dot{p}_x$ and $\dot{p}_y$ backward in time \cite{ElgartPRE70,Stepanov,LM2011}. Here we employed the shooting method and explored different parts of the parameter space $a$, $b$ and $\Gamma$.
Figure \ref{nodeinst} shows typical examples of the numerically found instantons $M_3F_1$ and $M_3F_2$ in the case when $M_3$ is a node. Figures \ref{focinst31} and \ref{focinst32} refer to the case when $M_3$ is a focus.
As one can see, the instanton $M_3F_2$ is qualitatively similar, in both regimes, to that found in the conventional SI model \cite{MeersonPRE77}. The instanton $M_3F_1$ is different, as it corresponds to extinction of both sub-populations, absent in the conventional SI model.

Note that Fig.~\ref{focinst32}a resembles Fig.~\ref{MCpp}a, and Fig.~\ref{focinst31}a resembles Fig.~\ref{MCpp}b. In particular,  the instanton $M_3F_2$ passes close to the point $(0,0)$. This explains
the feature observed, for the same set of parameters $a$, $b$ and $\Gamma$, in stochastic simulations, see Fig. \ref{MC}a.
As in other stochastic systems exhibiting rare large fluctuations, one should expect that by averaging over a
a large number of stochastic trajectories, conditional on a given extinction route, one will obtain the corresponding instanton up to an error that vanishes as $N\to \infty$.

\begin{figure}
\includegraphics[width=2.3 in,clip=]{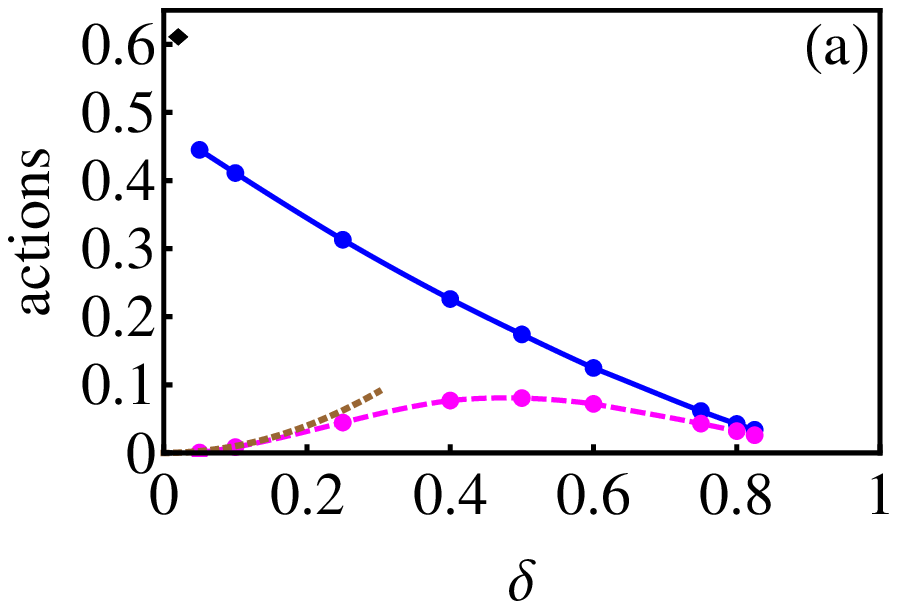}
\includegraphics[width=2.3 in,clip=]{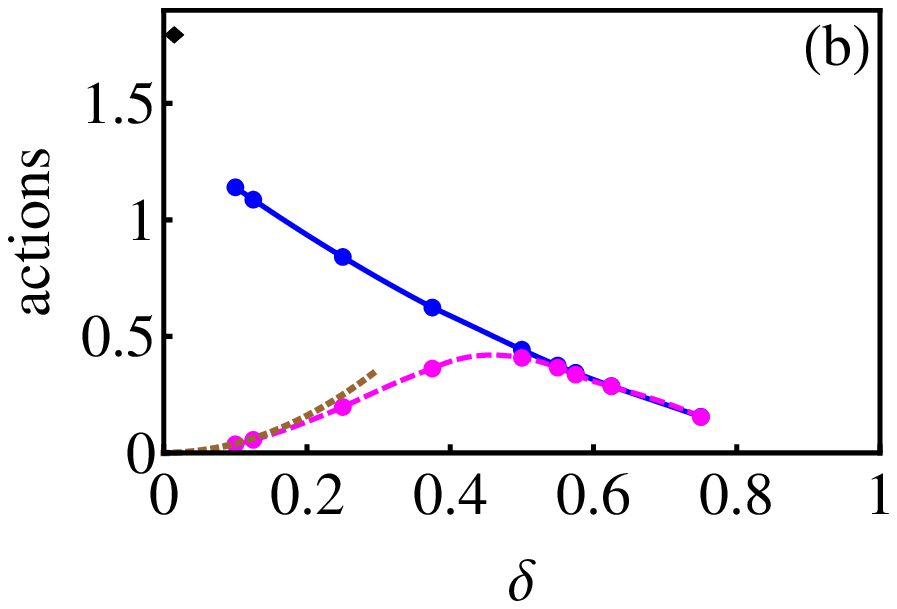}
\caption{(Color online) Accumulated actions $S_{31}$ (solid line) and $S_{32}$ (dashed line) as functions of $\delta=1-\Gamma/\Gamma_*$  for a fixed $b=1$ and two different values of $a$: $2$ (a) and $3$ (b). The dotted lines show the analytical asymptote from Eq.~(\ref{S32Analytic}). The diamonds denote the values of $S_{21}$ from Eq.~(\ref{S21}).}\label{Action}
\end{figure}

Figure \ref{Action}a shows the accumulated actions $S_{31}$ and $S_{32}$ for fixed $a=2$ and $b=1$ and different values of $\Gamma$ (parameterized by $\delta=1-\Gamma/\Gamma_*$).  The first observation is that  $S_{31}>S_{32}$ for all $\delta$, implying that the effective transition rate $r_{32}$ is \emph{exponentially} greater than $r_{31}$. As expected, $S_{32}$ vanishes at $\delta\to 0$ and  $\delta\to 1$ and has a maximum close to $\delta=1/2$, or $\Gamma=\Gamma_*/2$. This behavior follows that of the mean-field population size of foxes in the coexistence state. $S_{31}$ behaves differently: it has a maximum at $\delta\to 0$ (that is, at $\Gamma\to \Gamma_*$), goes down monotonically with an increase of $\delta$, and tends to zero as $\delta$ approaches $1$ (that is, $\Gamma$ approaches $0$). This behavior can be qualitatively understood from the $\Gamma$-dependence of the fixed points, see Eq.~(\ref{FixedPoints}). There is, however, one surprising feature here. As $\delta$ approaches zero, the fixed point $M_3$ becomes closer and closer to the fixed point $M_2$. One might expect, therefore,  that the instanton $M_3F_1$ and the accumulated action $S_{31}$ would approach the instanton $M_2F_1$ and the action $S_{21}$, respectively.  This is not what happens. Extrapolating the numerically calculated values of $S_{31}$ toward $\delta=0$, we obtain $S_{31}(\delta=0)\simeq 0.48$. This is considerably less than $S_{21}(\delta=0)=0.6137\dots$, obtained from Eq.~(\ref{S21}), see Fig.~\ref{Action}a. Furthermore, the numerically found instantons $M_3F_1$ at small $\delta$ are markedly different from the instanton $M_2F_1$ for which $y\equiv p_y \equiv 0$. Not only the instantons $M_3F_1$ exhibit relatively large intermediate values of $y$ and $p_y$, 
but the maximum value of $p_y$ along the instanton actually \emph{increases} as $\delta$ goes down. That is, as $\delta \to 0$, or $\Gamma \to \Gamma_*$, the fluctuations in the number of foxes play an important role in the joint extinction of the rabbits and foxes.

We also observed these features for other values of parameters $a$, $b$ and $\Gamma$ from the coexistence region $0<\Gamma<\Gamma_*$. For example, Fig. \ref{Action}b shows the $\delta$-dependence of the accumulated actions $S_{31}$ and $S_{32}$ for $a=3$ and $b=1$. Here the two actions
are very close to each other in a broader region of $\delta$, but the double inequality $S_{32}<S_{31}<S_{21}$ still holds.  Furthermore, it continues to hold when $a$ is only slightly greater than $b$ (not shown here).

When amplified by the very large factor $N\gg 1$ in the exponent, see Eq.~(\ref{rates}), the double inequality $S_{32}<S_{31}<S_{21}$ leads to
\begin{equation}\label{RateRelations}
    r_{21}\ll r_{31}\ll r_{32},
\end{equation}
implying that the sequential extinction route (foxes first, rabbits second) is much more likely than
the (almost) parallel extinction route. Using the inequality  $r_{32}\gg r_{31}$, we can further simplify
Eqs.~(\ref{P1long})-(\ref{MTEtotal}). In particular, the mean time to extinction of foxes becomes simply $\tau_F\simeq 1/r_{32}$, whereas the mean time to extinction of both sub-population is $\tau_{RF} \simeq 1/r_{21}$. Here the quantities $\tau_F$ and $\tau_{RF}$ are determined by the ``kinetic bottlenecks" of the effective transitions, as expected.

However, for a sufficiently high predation rate, and not too large $N$, the transition rates $r_{31}$ and $r_{32}$ are comparable, and Eqs.~(\ref{P1long})-(\ref{MTEtotal}) should be used in their complete form.

\subsubsection{Proximity of instantons $M_3 F_1$ and $M_3 F_2$ at small $\Gamma$}
\label{sec:closeness}
\begin{figure}
\includegraphics[width=2.5 in,clip=]{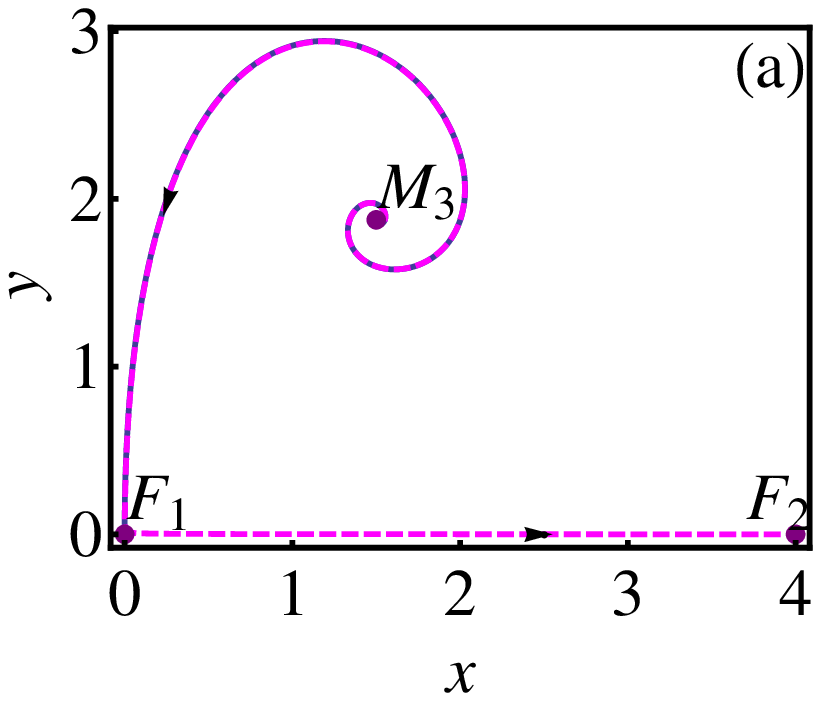}
\includegraphics[width=2.5 in,clip=]{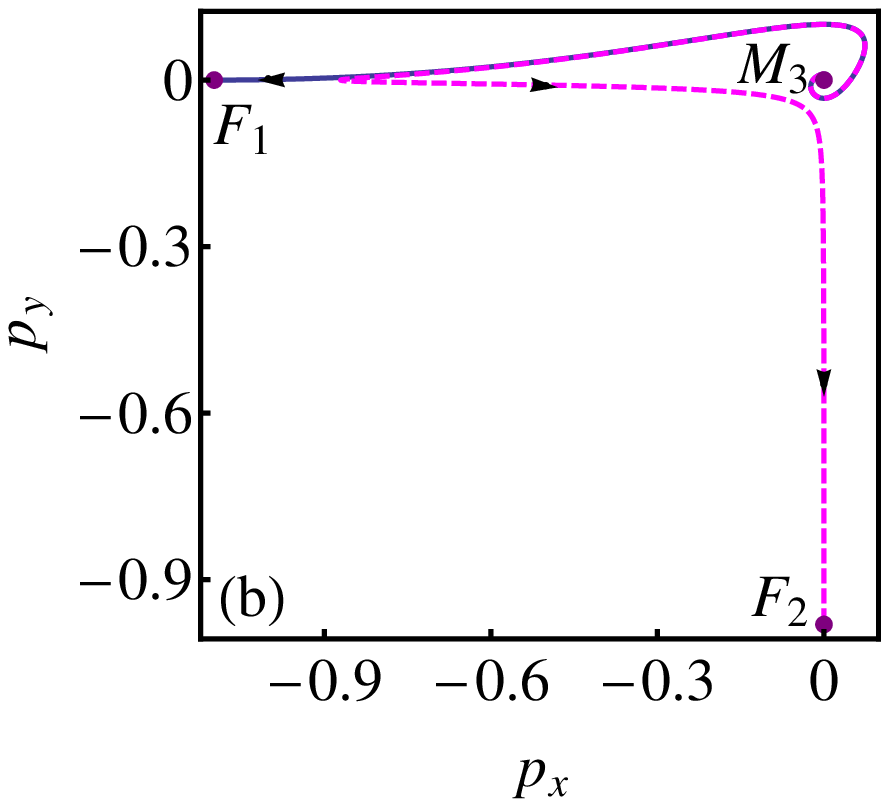}
\caption{(Color online) $xy$ (a) and $p_xp_y$ (b) projections of the instantons $M_3F_1$ (solid lines) and $M_3F_2$ (dashed lines) for $a=3,\,b=1$ and $\Gamma=1.5$. For these parameters $\delta=0.625$, and the actions $S_{31}$ and $S_{32}$ are very close, see Fig.~\ref{Action}b.}
\label{closeinst}
\end{figure}

As is evident from Fig.~\ref{Action}, the actions $S_{31}$ and $S_{32}$ approach each other closely for sufficiently large $\delta$, that is for high predation rates. The reason for it becomes clear upon comparison of the instantons  $M_3 F_1$ and $M_3 F_2$ for sufficiently small $\Gamma$, see  Fig.~\ref{closeinst}. One can see that the instanton $M_3F_2$ almost coincides with the instanton $M_3F_1$ until, close to the fixed point $F_1$,  it abruptly changes its direction and goes toward the fixed point $F_2$. On the latter segment of the trajectory the values of $y$ and $p_x$ stay close to zero, so the contribution of this segment to the action $S_{32}$ is negligible. In other words,
the most likely route to extinction of the foxes in this regime of parameters  involves a drastic decrease in the number of rabbits: almost all the way to their extinction! Then, a few remaining rabbits start reproducing, at a very small number of foxes, and the rabbit population gets reestablished. Note that, in the WKB formalism, the reestablishment of the rabbits, accompanied by extinction of the foxes, occurs via the ``fluctuational" fixed point $F_2$, rather than via directly  approaching the mean-field point $M_2$.

\subsubsection{Two modifications of the model}

One central result of this work is the strong inequality $r_{31}\ll r_{32}$ observed in most of the parameter space that we explored. That is, the sequential extinction route (first the predators then, after a long time, the prey) is usually much more likely than the (almost) parallel route. In other words, the predators are usually much more vulnerable extinction-wise than their prey. In retrospect, this feature is hardly surprising: As the predators are dependent on their prey for survival, they can be \emph{at best} as resilient extinction-wise as the prey. (This also gives an intuitive explanation to the proximity of the instantons $M_3F_2$ and $M_3F_1$ at high predation rates, reported in the previous subsection.) Will this feature hold for other predator-prey models? To get insight, we introduced two minor modifications of our model, each of them endowing the predators with more resilience  but still keeping them dependent on the prey for survival.  In one modification we added a direct reproduction process $F\rightarrow2F$ with rate $d<1$. This can model a large amount of additional small prey for the foxes, say mice.  The direct reproduction reduces the effective death rate of the foxes. Still, as long as $d<1$, the foxes go extinct deterministically in the absence of rabbits. In another modification of the model we replaced the predation process $F+R\rightarrow2F$ by a more efficient one $F+R\rightarrow3F$.  Our numerics showed that, for both of these modified models, the inequality $S_{3,1}>S_{3,2}$ continues to hold. Furthermore, in both models we observed, for sufficiently high predation rates, the convergence of the instantons $M_3F_2$ and $M_3F_1$ to each other until a close vicinity of the fixed point $F_1$ is reached. A difference between the original model and the modified ones is that, as we improve the conditions for the predators, this convergence occurs at lower predation rates.

\section{Conclusions}
\label{sec:discussion}

We considered a simple stochastic predator-prey model in its coexistence region and observed that this model exhibits two different routes of extinction of each of the populations: the sequential and the (almost) parallel.  This multiplicity of the extinction routes can be  conveniently accounted for by an effective three-state master equation for the probabilities to observe the coexistence state, the predator-free state and the empty state. The WKB approximation yields the effective transition rates between these three states, as well as the most likely paths of the sub-populations to extinction: the instantons. We showed numerically that the parallel extinction route is usually
much less likely than the sequential one, implying a great robustness of the prey against predation. For a sufficiently high predation rate, however, the two routes may have comparable probabilities.

The rest of parameters being fixed, our model predicts an optimal predation rate so that the mean time to extinction of the predators is maximum. Not surprisingly, this optimal predation rate is close to that for which the quasi-stationary population size of the predators is maximum. A surprising result is that, for sufficiently high predation rates, the optimal path to extinction of the predators almost coincides with the optimal path to the joint extinction of the predators and prey until a point (corresponding to an almost zero size of each sub-population) is reached where
the two optimal paths depart from each other: the predator population continues moving toward extinction whereas the prey population reestablishes itself. That is, for a high predation rate, the predators are more likely to reach extinction by consuming nearly all the prey. This phenomenon appears to be robust and independent of details of the predator-prey model, as long as predators still need prey for their survival.

Finally, all our results can be reformulated in terms of the SI epidemic model for an isolated community, see Table 2, where both the infected, and the susceptible populations are prone to extinction.

\section*{ACKNOWLEDGMENTS}

We acknowledge useful discussions with Michael Khasin and Pavel Sasorov. This work was supported by the Israel Science Foundation (Grant No. 408/08) and
by the US-Israel Binational Science Foundation (Grant No.
2008075).


\begin{thebibliography}{99}
\bibitem{Bartlett} M. S. Bartlett, \emph{Stochastic Population Models in Ecology and
Epidemiology} (Wiley, New York, 1961).
\bibitem{TREE} O. Ovaskainen and B. Meerson, Trends in Ecology and Evolution \textbf{25}, 643 (2010).
\bibitem{Turner} J. W. Turner and M. Malek-Mansour, Physica A 93, 517 (1978).
\bibitem{ElgartPRE70} V. Elgart and A. Kamenev, Phys. Rev. E \textbf{70}, 041106 (2004).
\bibitem{Doering} C.R. Doering, K.V. Sargsyan, and L.M. Sander, Multiscale Model. and Simul. \textbf{3}, 283 (2005).
\bibitem{AM2006} M. Assaf and B. Meerson, Phys. Rev. Lett. \textbf{97}, 200602 (2006); Phys. Rev. E \textbf{75}, 031122 (2007).
\bibitem{Frey} T. Reichenbach, M. Mobilia, and E. Frey, Phys. Rev. E \textbf{74}, 051907 (2006).
\bibitem{Kessler} D. A. Kessler and N. M. Shnerb, J. Stat. Phys. 127, 861 (2007).
\bibitem{plasmids} M. Assaf and B. Meerson, Phys. Rev. Lett. \textbf{100}, 058105 (2008).
\bibitem{DykmanPRL101} M. I. Dykman, I. B. Schwartz, and A. S. Landsman, Phys. Rev. Lett. \textbf{101}, 078101 (2008);
I. B. Schwartz, L. Billings, M. Dykman, and A. Landsman, J. Stat. Mech. P01005 (2009).
\bibitem{MeersonPRE77} A. Kamenev and B. Meerson, Phys. Rev. E \textbf{77}, 061107 (2008).
\bibitem{KMS} A. Kamenev, B. Meerson, and B. Shklovskii, Phys. Rev. Lett. \textbf{101}, 268103 (2008).
\bibitem{AKMmod} M. Assaf, A. Kamenev, and B. Meerson, Phys. Rev. E \textbf{78}, 041123 (2008).
\bibitem{MeersonPRE79} M. Assaf, A. Kamenev, and B. Meerson, Phys. Rev. E \textbf{79}, 011127 (2009).
\bibitem{KamPark} M. Parker and A. Kamenev, Phys. Rev. E \textbf{80}, 021129 (2009);
J. Stat. Phys. \textbf{141}, 201 (2010).
\bibitem{MSepidemic} B. Meerson and P.V. Sasorov, Phys. Rev. E \textbf{80}, 041130 (2009).
\bibitem{KhasinPRL103} M. Khasin and M. I. Dykman, Phys. Rev. Lett. \textbf{103}, 068101 (2009).
\bibitem{AssafPRE81} M. Assaf and B. Meerson, Phys. Rev. E \textbf{81}, 021116 (2010).
\bibitem{timeresolved} M. Khasin, B. Meerson, and P. V. Sasorov, Phys. Rev. E \textbf{81}, 031126 (2010).
\bibitem{MeersonPRE81} M. Khasin, M. I. Dykman, and B. Meerson, Phys. Rev. E \textbf{81}, 051925 (2010).
\bibitem{AMS} M. Assaf, B. Meerson, and P. V. Sasorov, J. Stat. Mech. P07018 (2010).
\bibitem{MS10} B. Meerson and P.V. Sasorov, Phys. Rev. E \textbf{83}, 011129  (2011).
\bibitem{KD2} M. Khasin and M. I. Dykman, Phys. Rev. E \textbf{83}, 031917 (2011).
\bibitem{LM2011} I. Lohmar and B. Meerson, Phys. Rev. E \textbf{84}, 051901 (2011).


\bibitem{Murray} A.J. Lotka, Proc. Natl. Acad. Sci. U.S.A. \textbf{6}, 410 (1920); V. Volterra,  \textit{Le\c{c}ons sur la Th\'{e}orie Math\'{e}matique de la Lutte Pour la Vie} (Gauthier-Villars, Paris, 1931); J. D. Murray, \emph{Mathematical Biology: I. An Introduction} (Springer, Berlin, 1989).
\bibitem{SI} H. Hethcote, Math. Biosci. \textbf{28}, 335 (1976); SIAM
Rev. \textbf{42}, 599 (2000); I. N{\aa}sell, J. R. Stat. Soc. Ser. B \textbf{61}, 309 (1999).
\bibitem{Grasman} O. A. van Herwaarden and J. Grasman, J. Math. Biol. \textbf{33}, 581 (1995).
\bibitem{vH} O. A. van Herwaarden, J. Math. Biol. \textbf{35}, 793 (1997).

\bibitem{Kubo} R. Kubo, K. Matsuo, and K. Kitahara, J. Stat. Phys. \textbf{9}, 51  (1973).
\bibitem{Hu} G. Hu, Phys. Rev. A \textbf{36}, 5782 (1987).
\bibitem{Peters} C.S. Peters, M. Mangel, and R. F. Costantino,  Bull. Math. Biol. \textbf{51}, 625 (1989).
\bibitem{DykmanPRE100} M. I. Dykman, E. Mori, J. Ross, and P. M. Hunt, J. Chem. Phys. \textbf{100}, 5735 (1994).



\bibitem{MSexplosion} B. Meerson and P. V. Sasorov, Phys. Rev. E \textbf{78}, 060103(R) (2008).
\bibitem{Escudero} C. Escudero and A. Kamenev, Phys. Rev. E \textbf{79}, 041149
(2009).


\bibitem{Kaplan} B. Meerson, P. V. Sasorov, and Y. Kaplan, Phys. Rev. E \textbf{84}, 011147  (2011).
\bibitem{MS11} B. Meerson and P.V. Sasorov, Phys. Rev. E \textbf{84}, 030101(R) (2011).

\bibitem{Assaf2011} M. Assaf, E. Roberts, and Z. Luthey-Schulten, Phys. Rev. Lett. \textbf{106}, 248102 (2011).
\bibitem{EK2} V. Elgart and A. Kamenev, Phys. Rev. E 74, 041101 (2006).
\bibitem{FW} M. I. Freidlin and A. D. Wentzell, \emph{Random Perturbations
of Dynamical Systems} (Springer-Verlag, New York, 1998), 2nd ed.

\bibitem{Stepanov} A. I. Chernykh and M. G. Stepanov, Phys. Rev. E \textbf{64}, 026306
(2001).


\end{thebibliography}
\end{document}